# Timing resistive plate chambers for thermal neutron detection with 3D position sensitivity

L.M.S. Margato[1,*], G. Canezin[1], A. Morozov[1], A. Blanco[1], J. Saraiva[1], L. Lopes[1], P. Fonte[1,2]

[1] LIP-Coimbra, Departamento de Física, Universidade de Coimbra, 3004-516 Coimbra, Portugal

[2] ISEC - Instituto Superior de Engenharia de Coimbra, 3031-199 Coimbra, Portugal

**Abstract**: An optimized design of a neutron detector based on timing RPCs (Resistive Plate Chambers) with boron-10 neutron converters is presented. The detector is composed of a stack of ten double gap RPCs with aluminium cathode plates coated on both sides with $^{10}B_4C$. This design enables simultaneous determination with high accuracy of both the neutron time-of-flight (down to ns resolution) and the interaction position in 3D (down to 0.25 mm resolution across and ~1 mm along the beam). It is shown that the detection efficiency can approach 60% for neutrons with lambda = 4.7 Å. A new geometry with less material budget is introduced for the signal pick-up strip arrays. The results of simulation-based optimization of the design are reported considering the trade-off between the detection efficiency, the count rate capability and the amount of elastic scattering on the detector components.

## 1. INTRODUCTION

Modern instruments for neutron science applications, such as small-angle neutron scattering, reflectometry and macromolecular crystallography require thermal neutron detectors with high detection efficiency, low sensitivity to gamma rays, high counting rate capability and high spatial resolution [1,2]. In many applications, especially at neutron spallation sources, the detector must also be able to read-out each neutron's time-of-flight (ToF) to enable wavelength-resolved measurements. Recently, we have introduced position-sensitive thermal neutron detectors based on double gap RPCs (Resistive Plate Chambers) combined with $^{10}B_4C$ solid neutron converters. It was already demonstrated that such detectors can provide detection efficiency above 50% and sub-millimeter spatial resolution [3]. RPCs are also well known for their fast response (sub-ns range) [4], making them excellent for timing purposes. Based on the $^{10}B$-RPC neutron detection technology, we discuss here a concept of a timing RPC-based neutron detector, with 3D position readout (XYZ) and timing capability. This type of detector is intended for ToF neutron diffraction and reflectometry, wavelength resolved neutron imaging and other applications requiring simultaneous readout of position and time. We also report the results of a Geant4 simulation-based optimization study considering the trade-off between the detection efficiency, the maximum count rate and the amount of elastic neutron scattering on the detector components.

## 2. TIMING RPC NEUTRON DETECTOR

The design of the detector proposed here consists of a stack of double gap RPCs, oriented normally to the direction of the incident neutrons. Each RPC consists of two resistive anodes and an aluminum cathode between them, all parallel to each other. The electrodes are separated by a 0.28 mm wide gas gap. The cathode plate is lined on both sides with a ~1 μm thick layer of $^{10}B_4C$. The deposition can be performed using the magnetron sputtering technique which, as has already been demonstrated, can provide stable adhesion and good radiation hardness [5,6]. A schematic drawing of the detector layout is shown in Fig. 1. For a detailed description of the RPC components see [7]. As a standalone double gap RPC provides less than 10% detection efficiency for thermal neutrons [3], stacking of several RPCs is required to achieve high detection efficiencies. Such stacking of RPCs also improves the count rate capability (see below).

The cathode of each RPC is read individually. In this way, the position of the neutron capture along

---

** Corresponding author.

E-mail address: margato@coimbra.lip.pt (L.M.S. Margato)

the stack (Z coordinate) and the ToF can be determined using the fast electronic component of the cathode signal.

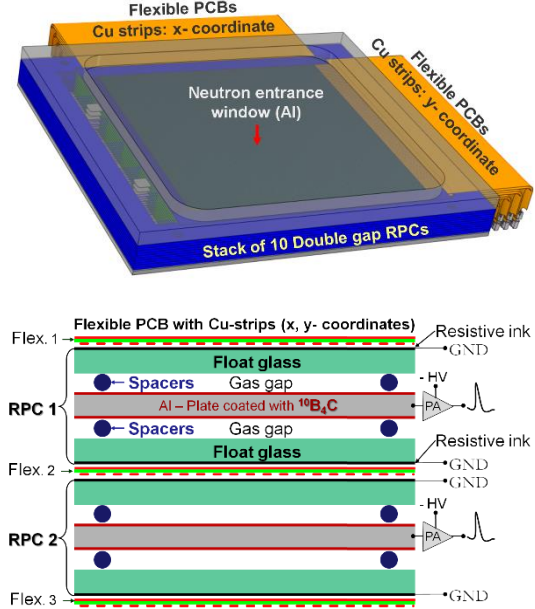

Fig. 1. At the top is a 3D view of the prototype, consisting of a stack of 10 double gap RPCs coated with $^{10}B_4C$. At the bottom is a schematic drawing (dimensions are not to scale) illustrating the RPCs design, and how the stack is formed (only two RPCs are shown).

Benefiting from the fast timing (sub-ns) of RPCs and taking into account ~1 ns flight time of thermal neutrons through the $^{10}B_4C$ layer of about 1 μm thick, the ToF can be measured up to very high accuracy. Usually, the start trigger for ToF is defined by the choppers or by the pulse of the spallation source accelerator.

Arrays of signal pick-up strips are used to determine the XY coordinates of the neutron events. A thin polyimide film (25 μm thick) holding on each side an array of parallel metallic strips (25 μm thick copper, 1 mm pitch, 0.4 mm wide), orthogonal to each other, is inserted in between each neighboring RPC, as well as before the first and after the last RPC of the stack. In this arrangement, the gas gaps of adjacent RPCs share the same arrays of strips. For X, as well as Y coordinates, the individual strips of each array with the same index are interconnected and readout with the same electronic channel. This approach allows to reduce the number of channels by a factor of N+1, where N is the number of double-gap RPCs in the stack. Note that a neutron detection event produces ionization only in one gas gap. Therefore, once the cathode with a signal is identified, the triggered double gap RPC in the stack is known. However, it remains undefined in which side of the cathode of the respective RPC the neutron capture has occurred (see Fig. 1). This ambiguity can be resolved by comparing the sum signal for X and Y strips if all signal pick-up arrays are installed identically over the stack: for example, the X-coordinate strips are all situated at the upper side of the polyimide films (see Fig. 1). In this configuration, for an event in the upper gas gap of any double gap RPC, the X strips are partially screened by the Y strips, and, therefore, the sum Y signal is larger. Contrary, for each lower gas gap event, the sum signal of X strips is larger than the one for Y strips. This signal pickup strips design is better considering the material budget with respect to the one used in the previous prototype [3], where two polyimide films and three arrays of strips were used per each flexible PCB.

3. DETECTOR DESIGN OPTIMIZATION

The simulation model of the detector used in this optimization study consists of a stack of 10 double gap RPCs (20 layers of $^{10}B_4C$) and 11 sets of XY signal pickup units. This number of RPCs was chosen to achieve detection efficiencies of about 50% for thermal neutrons. Note that the number of RPCs should be selected for each particular application, considering the neutron wavelength and the specific efficiency and counting rate requirements.

The RPCs components have the following properties:

- 0.33 mm thick soda lime glass anodes;
- 0.3 mm thick cathodes made of 5754 aluminium alloy;
- $^{10}B_4C$ converters ($^{10}B$ enrichment level of ~97%) of various thickness;
- 0.28 mm wide gas gaps filled with R134a gas at atmospheric pressure.

The anode and cathode thicknesses were reduced from the value of 0.5 mm used in the previous prototype [3] in order to decrease the elastic neutron scattering. The new values are close to the limit defined by the mechanical properties. It was assumed that the flexible PCBs for the position readout had a 25 μm thick polyimide film with 18 μm thick copper layers on each side.

All simulations were performed using the Geant4 toolkit v10.7.2 [8] with ANTS2 v4.36 [9] as front-end. High precision neutron physics was activated by selecting the QGSP-BIC-HP reference physics list. The primary neutrons were generated as a monochromatic (4.7 Å) pencil beam with normal incidence at the center of the detector.

Typically, simulations were performed with $10^6$ primary neutrons. A neutron was considered detected if the capture reaction products deposited at least 100 keV in the gas gap. This threshold value was selected as the one giving a good match between the simulated and experimentally observed detection efficiency [10].

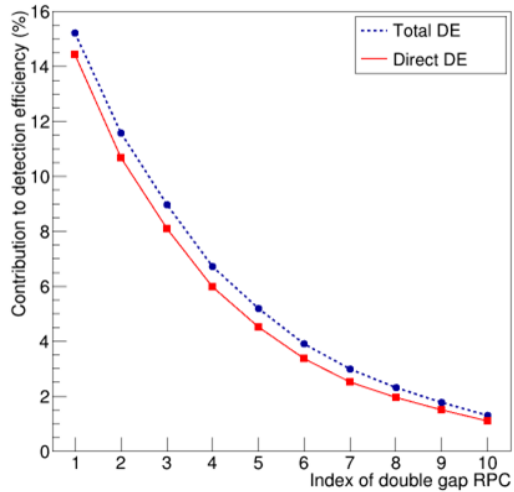

Fig. 2. Contribution to the overall detection efficiency of each double gap RPC in the stack, with $^{10}B_4C$ layers of the same thicknesses (1.15 μm). The sum of all contributions adds up to a detection efficiency of 60.4%. The "direct DE" data show the detection efficiency excluding those neutrons which had prior elastic scattering. The lines are an eye-guide connecting the data points.

As a starting point for the optimization, a detector with all $^{10}B_4C$ layers of the same thickness of 1.15 μm was considered. The overall detection efficiency was found to be 60.4 %. Fig. 2 shows the contributions to this efficiency from individual double gap RPCs. Since each next RPC down the beam direction (larger RPC indexes) "sees" less neutrons than the previous one by the same factor, the contribution reduces exponentially with the index. The figure also shows the "direct" detection efficiencies, which take into consideration only those neutrons which had no scattering interactions before detection. While such neutron events cannot be identified in a practical detector, this approach helps to estimate the fraction of the detected events which carry distorted spatial information and thus contribute only to the background.

The detection efficiency data of Fig. 2 also describe the relative counting rates of the individual RPCs. RPC1 has the highest contribution to the detection efficiency and thus should reach its maximum counting rate limit at a significantly lower neutron flux compared to, e.g., RPC10. Thus, this RPC establishes the limit on the detector response linearity with the flux.

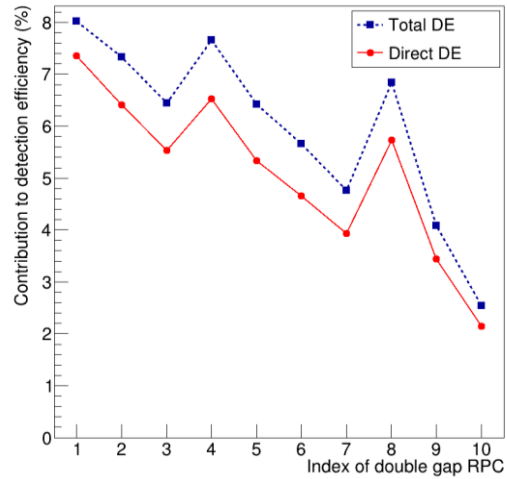

Fig. 3. Contribution to the detection efficiency of each double gap RPC in the stack for the configuration with optimized thicknesses. The sum of all contributions adds up to 59.8% total detection efficiency. The "direct DE" data show the detection efficiency excluding those neutrons which had prior elastic scattering. The lines are an eye-guide connecting the data points.

These considerations suggest that the detector design should be optimized using a procedure which not only maximizes the direct detection efficiency, but also attempts to make as equal as possible the contributions to the counting rate from all double-gap RPCs. The main parameter that affects the relative detection efficiencies of individual RPCs is the thickness of the converter layer. Due to the practical considerations related to the manufacturing of the $^{10}B_4C$ converters, we have decided to use just 3 different thicknesses: one for RPCs indexes from 1 to 3, another for indexes from 4 to 7 and the third for the ones from 8 to 10.

This conditional optimization was performed using the procedure described in [8] (the cost function is determined by both the overall direct detection efficiency and the equality factor of the contributions from the individual RPCs) to find the three optimal thickness values. The following values were obtained:

- 0.4 μm for RPC 1 to 3,
- 0.6 μm for RPC 4 to 7 and
- 2.2 μm for RPC 8 to 10.

Fig. 3 shows the individual contributions of each RPC to the detection efficiency, computed using the optimized layer thicknesses. The total detection efficiency (all RPCs together) is 59.8%, with an overall direct efficiency of 51.1%. The figure also shows that the differences in the contributions from the individual RPCs are much smaller compared to the configuration with equal thicknesses (see Fig. 2). The corresponding equality parameter (see [8] for the mathematical definition) increase by a factor of about two: from 0.38 to 0.70. Therefore, the optimized configuration should allow a factor of two higher neutron flux without reaching saturation. The results also show that for this configuration, in comparison to a standalone double-gap RPC, the linear regime is extended by 7 times (or by 14 times compared to a single gap RPC).

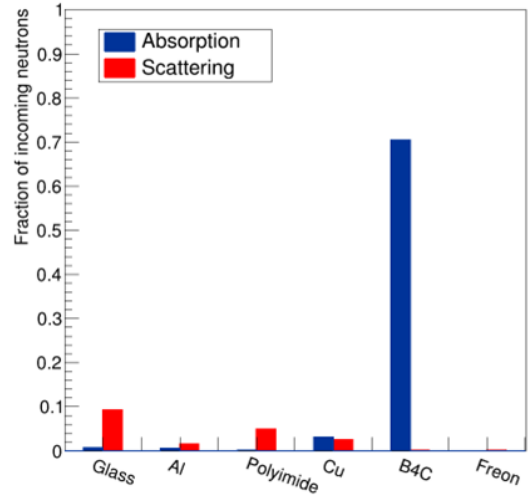

Fig. 5. Fractions of the primary neutrons in the beam lost due to absorption and elastic scattering in all materials of the detector components.

Fig. 5 shows simulation results giving the fractions of the primary neutrons in the beam lost due to absorption and elastic scattering on all materials of the detector components. All captures, except in $^{10}B_4C$, lead to a reduction of the detection efficiency. All scatters, if followed by a neutron detection, contribute to the neutron background. The strongest scattering is on glass, but it is extremely difficult to reduce the glass thickness any further. The polyimide and copper also have a significant contribution to scattering, but their effect is noticeably decreased due to the use of a new configuration of the readout arrays.

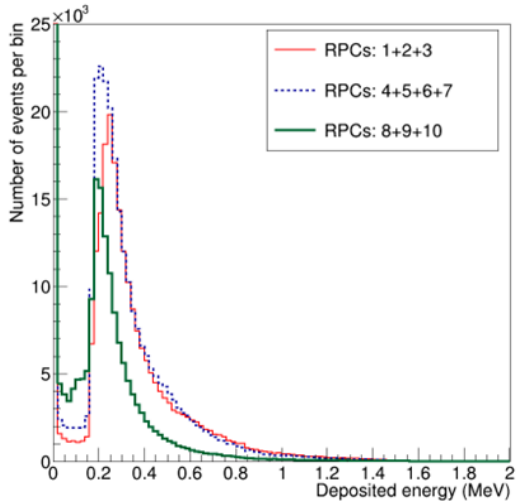

Fig. 4. Distribution of energy deposited in the gas gap, simulated with Geant4 for RPCs with different $^{10}B_4C$ thickness: RPC 1 to 3 of 0.4 μm, RPC 4 to 7 of 0.6 μm and RPC 8 to 10 of 2.2 μm.

For the optimized configuration we have also computed the distribution of the energy deposited in the gas gap by $^4$He and $^7$Li particles originating from the $^{10}B(n, α)^7Li$ neutron capture reaction. The results for RPCs with three different thickness are shown in Fig. 4. The analysis of the results shows that all energy deposition distributions exhibit a narrow profile. This is explained by a very narrow (0.28 mm) width of the gas gap, limiting the range of $^4$He and $^7$Li particles escaping from the $^{10}B_4C$ layer to the gas. The width of the profile for the RPCs 8, 9 and 10 is significantly narrower. In this case the reaction products have to travel, on average, a longer distance inside the $^{10}B_4C$ layer and thus enter the gas gap with less energy compared to the cases with thinner converter layers. However, for all profiles, the "leading edge" before the peak appears essentially at the same energy value (~170 keV). Thus, the same energy threshold can be used for all three groups of RPCs without a significant impact on the detection efficiency.

## 4. CONCLUSIONS

A detector design for thermal neutrons with 3D position sensitivity and timing capability is described. A simulation-based optimization study suggests that, using only three different $^{10}B_4C$ thicknesses, it is possible to construct a detector with a total detection efficiency of about 60% (4.7 Å), and a quite flat distribution of the contributions from the individual RPCs to the efficiency. The latter property is important as it increases the range in neutron flux for which the detector is capable to operate in a linear regime. A new configuration for the arrays of readout strips is introduced, allowing to reduce the material budget of the detector and leading, in turn, to a decrease in the neutron scattering background. A detector prototype based on the presented concept is being assembled, and a test in a neutron beam is already scheduled.

ACKNOWLEDGMENTS

This work was supported by Portuguese national funds OE and FCT-Portugal (grant EXPL/FIS-NUC/0538/2021).